\documentclass[journal]{IEEEtran}
\ifCLASSINFOpdf
\else
\fi
\usepackage[cmex10]{amsmath}
\usepackage{amsfonts}
\usepackage{amsthm}
\usepackage{amssymb}
\usepackage{latexsym}
\usepackage[latin1]{inputenc}

\begin{document}

\newtheorem{theo}{Theorem}[section]
\newtheorem{lemma}[theo]{Lemma}
\newtheorem{cor}[theo]{Corollary}

\newtheorem{definition}[theo]{Definition}
\newtheorem{example}[theo]{Example}
\newtheorem{remark}[theo]{Remark}
\newtheorem{conj}[theo]{Conjecture}
\newtheorem{prop}[theo]{Proposition}
\newtheorem{reference}{}
\newcommand{\wh}{\widehat}
\newcommand{\ol}{\overline}

\newcommand{\Hom}{{\rm Hom}\,}           
\newcommand{\Aut}{{\rm Aut}}             
\newcommand{\zzz}{\mathbb{Z}}        
\newcommand{\car}{{\rm char}}
\def\F{{\mathbb F}}
\newcommand{\N}{{\mathbb N}}              
\def\lcm{{\rm lcm}}

%
\title{Minimal Binary Abelian Codes of length $p^mq^n$}
%
%
%

\author{Gladys~Chalom, Raul~Antonio~Ferraz,
        Marinês Guerreiro,
        and~César~Polcino~Milies,
\thanks{G. Chalom, R.A. Ferraz and C. Polcino Milies are with the Instituto de Matemática e Estatística, Universidade de São Paulo, Caixa Postal 66281, CEP 05314-970 - São Paulo - SP (Brasil), supported by CNPq, Proc. 300243/79-0(RN) and FAPESP, Proc. 09/52665-0. E-mail: agchalom@ime.usp.br, raul@ime.usp.br and polcino@ime.usp.br.}
\thanks{M. Guerreiro is with Departamento de Matemática, Universidade Federal de Vi\c cosa, CEP 36570-000 - Vi\c cosa-MG (Brasil), supported by 
CAPES, PROCAD 915/2010 and FAPEMIG, APQ CEX 00438-08. E-mail: marines@ufv.br.}
\thanks{This work was presented in part at the conference Groups, Rings and Group Rings, July 2012, Ubatuba-SP. }
\thanks{Manuscript received ???, 20012; revised ???, 2012.}}

%
%

\markboth{IEEE Transactions of Information Theory,~Vol.~??, No.~??, January~20??}%
{Shell \MakeLowercase{\textit{et al.}}: Bare Demo of IEEEtran.cls for Journals}
%



\maketitle

\begin{abstract}
We consider binary abelian codes of length $p^nq^m$ where $p$ and $q$ are prime rational integers under  some restrictive hypotheses. In this case, we determine the idempotents generating minimal codes and  either the respective weights or bounds of these weights. We give examples showing that these bounds are attained in some cases.
\end{abstract}


%
\IEEEpeerreviewmaketitle

\section{Introduction}
%
%
%
%
\IEEEPARstart{G}{enerators} of minimal abelian codes were 
determined in~\cite{FM} together with their corresponding dimensions and weights
under the following hypotheses: 
\begin{eqnarray}\label{fm}
& G &  \mbox{ a
finite abelian group of exponent } p^m  \nonumber
\\ & & \mbox{ (or $2p^m$, with
$p$ odd)}   \\
& \F & \mbox{  a field with $q$ elements}, \nonumber \\ & & q  \mbox{ with multiplicative order } \varphi(p^m) \!\!\!\mod p^m.  \nonumber
\end{eqnarray}

Here we extend these ideas  to study groups of the form $G_p\times G_q$, where $G_p$ and $G_q$ denote abelian groups, the first a  $p$-group and the second a $q$-group,
 satisfying the following conditions which will allow us to use the results in~\cite{FM}.
We shall denote by $U(\zzz_{n})$ the group of units of $\zzz_{n}$. 
\begin{eqnarray} \label{hypopq}
& (i) & \gcd (p-1,\,q-1 ) =  2, \nonumber \\
& (ii)  &  \  \bar{2}\;
\mbox{{\em generates the groups}}\; U(\zzz_{p^2})  \mbox{ {\em and} } 
\;U(\zzz_{q^2}) \\
& (iii)  & \gcd(p-1,q) = \gcd(p,q-1)=1. \nonumber
\end{eqnarray}

Notice that hypothesis $(i)$ above implies that at least one of the primes $p$ and $q$ is congruent to 3 (mod 4). In what follows, to fix notations, we shall always assume $q\equiv 3 \ $ (mod 4). 

Let $\,\F_2\,$ be the field with $2$ elements.
In Section~\ref{idemp} we compute, for  each minimal code in $\F_2(G_p\times G_q)$,  the  generating primitive idempotent,
the dimension and give explicitely a basis over $\F_2$. 

Primitive idempotents are usually determined using the characters of the group over a splitting field $\,L\,$ and
then using Galois descent
(see~\cite[Lemma 9.18]{isaacs}).
In what follows, we avoid the use of this technique by deriving the expression of the primitive idempotents from the subgroup structure of $G$.

In the later sections we show how to study the corresponding code weights in several cases.

\section{Basic Facts}\label{basic}

\hspace{.5cm} Let $\F_p$ be the Galois field with $p$ elements. In
this section we list some results on finite fields and elementary number theory that will be needed in the sequel. Our first result is well-known.

\begin{lemma}\label{lemars}
Let $p$ be a positive prime number and $\,r,\,s\in \N^*$. Then
$$\,\F_{p^r}\otimes_{\F_{p}} \F_{p^s}\,\cong
\,\gcd ( r,s)\cdot \F_{p^{\lcm (r,s)}}.$$
\end{lemma}

\begin{remark}\label{rem1}
Notice that any extension $L$ of $\,\F_2\,$ of even degree
contains a subfield $K$ with four elements, hence there exists an element $1\neq a\in L$ such that $\,a^3=1$.
\end{remark}

\begin{lemma}\label{lemae1e2}
Let $r,\,s\in{\mathbb N}$ be non-zero numbers such that $\gcd(r,s)=2$. Let $u\in \F_{2^{r}}\,$ and $v\in
\F_{2^{s}}\,$ be elements  satisfying the equation $\,x^2+x+1=0$.
Then
\begin{equation}\label{compuv}
\F_{2^{r}}\otimes_{\F_{2}}\,\F_{2^{s}}\cong \F_{2^{\frac{rs}{2}}}\oplus\,\F_{2^{\frac{rs}{2}}}
\end{equation}
and $\,e_1\,=\,(u\otimes v)\,+\,(u^2\otimes v^2)\,$
and $\,e_2\,=\,(u\otimes v^2)\,+\,(u^2\otimes v)\,$ are the primitive idempotents generating the simple components of~(\ref{compuv}).
\end{lemma}

\begin{IEEEproof} The decomposition of $\F_{2^{r}}\otimes_{\F_{2}}\,\F_{2^{s}}$ as a direct sum follows  from Lemma~\ref{lemars}.

Since $u,u^2\in \F_{2^{r}}$ (resp. $v,v^2\in \F_{2^{s}}$)
are linearly independent over $\F_{2}$, we have that $(u\otimes v),\,(u^2\otimes v^2),\,(u\otimes v^2)\,$ and $\,(u^2\otimes v)$ are linearly independent in $\displaystyle
\F_{2^{r}}\otimes_{\F_{2}}\,\F_{2^{s}}$. Hence  $e_1\neq 0$ and $e_2\neq 0$.
As $1+v+v^2=0$, $1+u+u^2=0$, and also $u^3=v^3=1,$ we obtain:
\begin{eqnarray*}
e_1\cdot e_2 & = & (u^2\otimes 1)+(1\otimes v^2)+(1\otimes v)+(u\otimes 1) \\
& = & ( u^2+u)\otimes 1 + 1\otimes (v+v^2)=0
\end{eqnarray*}
and also
\begin{eqnarray*}e_1+e_2 & = & (u\otimes v)+(u^2\otimes v^2)+(u\otimes v^2)+(u^2\otimes
v) \\ 
& = & u\otimes (v+v^2)+u^2\otimes(v+v^2)= 1\otimes 1.
\end{eqnarray*}

 As
$\,\F_{2^{\frac{rs}{2}}}\oplus\,\F_{2^{\frac{rs}{2}}}\,$ has two simple components, $\,e_1\,$ and $\,e_2\,$ are, in fact, the corresponding
primitive idempotents.
\end{IEEEproof}

We shall also need the following result whose proof is elementary.

\begin{lemma}\label{order2modpq}
Let $p$ and $q$ be two distinct odd primes such that $\gcd (p-1,\,q-1)=2$ and $\bar{2}$
generates both groups of units $U(\zzz_{p})$ and $U(\zzz_{q})$. Then the least positive integer $k$ such that
$2^k\equiv 1 (\!\!\!\mod pq)$ is $\lcm (p-1,q-1)= \frac{(p-1)(q-1)}{2}$.
\end{lemma}

\begin{remark} \label{6conjuntos} Let $p$ be an odd prime.
Denote by $Q$ the set of non-zero elements in $\zzz_p$ which are quadratic residues modulo $p$ and by
$N$  the set of non-zero elements in $\zzz_p$ which are non-quadratic residues modulo $p$.
By  \cite[Theorem 79]{Landau}, we have
$\zzz_p=\{0\} \cup Q\cup N$, hence $|Q|=|N|=\frac{p-1}{2}$.

We note that $\zzz_p$ is the union of the following six disjoint subsets:
\begin{eqnarray}\label{partitionZp6sets}
\{0\}, &  & \{1\},\nonumber \\
Q_Q &=& \{x\in \zzz_p : x \in Q \;\ \mbox{and}\;\ x-1\in Q\},\nonumber\\
Q_N & = & \{x\in \zzz_p : x \in Q \;\ \mbox{and}\;\ x-1\in N\}, \\
N_Q& = &\{x\in \zzz_p : x \in N \;\ \mbox{and}\;\ x-1\in Q\},\nonumber\\
N_N & = & \{x\in \zzz_p : x \in N \;\ \mbox{and}\;\ x-1\in N\}.\nonumber
\end{eqnarray}

For any odd prime $p$, clearly
$\bar{1}\in Q$ and if $\bar{2}$ generates $U(\zzz_p)$, then
$\bar{2}\in N$. As $\bar{1}\in Q$, we have $Q=\{\bar{1}\}\cup Q_Q\cup Q_N$ and $N=N_Q\cup N_N$.
\end{remark}

\begin{lemma} {\em (\cite[Theorem 83]{Landau})} \label{-1rq} Let $p$ be an odd prime.
The element $-\bar{1}$ is a quadratic residue in $\zzz_p$ if and only if \linebreak $p\equiv 1 (\!\!\!\mod 4)$.
\end{lemma}

The next elementary result is known and appears as Exercise 18, p. 149 in~\cite{NZM}.

\begin{lemma}\label{sizesets}
(a) Let $p\equiv 3 (\!\!\!\mod 4)$. Then $|Q_Q|=|Q_N|=|N_N|=\frac{p-3}{4}$ and $|N_Q|=\frac{p+1}{4}$.

\

(b) Let $p\equiv 1 (\!\!\!\mod 4)$. Then $|N_Q|=|Q_N|=|N_N|=\frac{p-1}{4}$ and $|Q_Q|=\frac{p-5}{4}$.
\end{lemma}

\section{Primitive Idempotents}\label{idemp}

\hspace{.5cm} 
Let $G$ be a group. 
For a subgroup $H\leq G$, we set $\widehat{H}=\displaystyle\sum_{h\in H}h$ and, for an element $x\in G$, we set $\widehat{x}=\wh{\langle x\rangle}$.

It was shown in ~\cite{FM} that, under the hypotheses in (\ref{fm}), if $A$ is an abelian $p$-group, then the primitive idempotents of $A$ can be constructed as follows: for each subgroup $H$ of $A$ such that $A/H \neq {1}$ is cyclic, consider the unique
subgroup $H^*$ of $A$ containing $H$ such that $|H^*/H|=p$.
We define $e_H=\widehat{H}-\widehat{H^*}$. Then, these elements $e_H$,  together with $e_A =\wh{A}$ are precisely the primitive idempotents of $FA$.

Moreover, if $[A:H]=p^r$ then the dimension of the ideal generated by an idempotent of the form $e_H$ is  
\begin{equation}\label{equation}
dim_\F(\F A)e_H = dim_\F\F[A/H]-dim_\F\F[A/H^*] = p^{r-1}(p-1)
\end{equation}
(see \cite[formula(1),p. 390]{FM}).
\\

Let $G_p$ and $G_q$ be finite abelian groups, the first a  $p$-group and the second a  $q$-group and set $G=G_p\times G_q$. For each subgroup $H$ in $G_p$ such that $G_p/H\neq 1$ is cyclic, consider the idempotent $e_H =\wh{H} -\wh{H^*}$ as above and, similarly, consider primitive idempotentes of the form $e_K = \wh{K} -\wh{K^*}$ for $G_q$.

Clearly $\wh{G_p}\cdot \wh{G_q}=\wh{G_p\times G_q}$ is a primitive idempotent of $\F_2 (G_p\times G_q)$.  

We claim that  idempotents of the form $\wh{G_p} \cdot e_K$ are primitive. In fact, we have $(\F_2G) \wh{G_p}\cdot e_K = (\F_2G \cdot \wh{G_p})e_K \cong (\F_2 G_q) e_K$ which is a field. In a similar way, it follows that idempotents of the form $e_H\cdot \wh{G_q}$ are primitive.

Finally, we wish to show that an idempotent of the form $e_H\cdot e_K$ decomposes as the sum of two primitive idempotents in $\F_2 G$.

To do so, write $e_H=\wh{H}-\wh{H^*}$ and set $a\in H^*\setminus H$. Notice that $aH$ is a generator of $H^*/H$. Set
\begin{equation}\label{uf2p}
u=\left\{
\begin{array}{ll}
a^{2^0}+a^{2^2}+\cdots +a^{2^{p-3}}, &
\mbox{if} \;p\equiv 1 (\!\!\!\!\!\mod 4)\;\mbox{or}\\
1+a^{2^0}+a^{2^2}+\cdots+a^{2^{p-3}},
& \mbox{if} \;p\equiv 3 (\!\!\!\!\!\mod 4) \end{array}\right.
\end{equation}
and
\begin{equation}\label{vf2p}
v=\left\{
\begin{array}{ll}
a^{2}+a^{2^3}+\cdots+a^{2^{p-2}}, &
\mbox{if} \;p\equiv 1 (\!\!\!\!\!\mod 4)\;\mbox{or}\\
1+a^{2}+a^{2^3}+\cdots+a^{2^{p-2}},
& \mbox{if} \;p\equiv 3 (\!\!\!\!\!\mod 4) \end{array}\right.
\end{equation}

Setting $(u\wh{H})^2 = u'\wh{H}$, a direct computation shows $(u'\wh{H})^2 = u\wh{H}$ and  $(u\wh{H})^2 +  u\wh{H} + e_H = 0$ (recall that the unity of $(\F_2G_p) e_H$ is precisely $e_H$). 

Also we have 
$$u\wh{H}(\wh{H}-\wh{H^*}) = u\wh{H} -u\wh{H^*} = u\wh{H} -\varepsilon(u)\wh{H^*}, $$
where $\varepsilon(u)$ is the number of summands in $u$, which is always even, hence $\varepsilon(u)=0$ and so  $u\wh{H}(\wh{H}-\wh{H^*})= u\wh{H}$. Similarly, we also have $u'\wh{H}(\wh{H}-\wh{H^*})= u'\wh{H}$.  Consequently, we see that both  $u\wh{H}$ and $u'\wh{H}$ lie in $\F_2G_p(\wh{H}-\wh{H^*})$.

Similarly, for $e_k=\wh{K}-\wh{K^*}$, set $b\in K^*\setminus K$ and define $v$ as in~\eqref{uf2p} and $v'$ as in~\eqref{vf2p} (replacing $a$ by $b$). Hence $v\wh{K}, \; v'\wh{K}=(v\wh{K})^2\in\F_2G_q(\wh{K}-\wh{K^*})$.

Notice that,  by equation~(\ref{equation}), we have $dim_\F (\F_2G_p) e_H = p^{r-1}(p-1)$ and $dim_\F (\F_2G_q) e_K = q^{s-1}(q-1)$, where 
\linebreak $r=[G_p:H]$ and $s=[G_q:K]$. Thus
$$(\F_2G_p) e_H \cong \F_{2^{p^{r-1}(p-1)}} \ \mbox { and } \  (\F_2G_q) e_K \cong \F_{2^{q^{s-1}(q-1)}}.$$

As $\gcd(p^{r-1}(p-1),q^{s-1}(q-1))=2$ we can apply Lemma~\ref{order2modpq} to see that 
\begin{eqnarray*}
e_1 = e_1(H,K) & = & u\wh{H}\cdot v\wh{K} \;+\; u'\wh{H}\cdot v'\wh{K} \\
e_2 = e_2(H,K) & = & u\wh{H}\cdot v'\wh{K} \;+\; u'\wh{H}\cdot v\wh{K}
\end{eqnarray*}
are primitive orthogonal idempotents such that $e_1+e_2 = e_He_K$.

Hence, we have shown the following.

\begin{theo}\label{teorema}
Let $G_p$ and $G_q$ be abelian $p$ and $q$-groups, respectively, satisfying the conditions in (\ref{hypopq}). For a group $G$, denote by $S(G)$ the set of subgroups $N$ of $G$ such that \linebreak $G/N\neq 1$ is cyclic. Then the set of primitive idempotents in $\F_2 [G_p\times G_q]$ is:
\begin{eqnarray*}
\wh{G_p} \cdot \wh{G_q}, && \\
  \wh{G_p}\cdot e_K,& &  K\in S(G_q),\\
 e_H\cdot \wh{G_q},& &  H\in S(G_p),\\
 e_1(H,K), \ e_2(H,K),& &   H\in S(G_p), K\in S(G_q).
 \end{eqnarray*}
\end{theo}

In the following sections we shall use this result to study minimal codes in some special cases and, in these cases, we shall also study the corresponding  weights and give explicit bases for these minimal codes.

\section{Codes in $\F_2 (C_p\times C_q)$}

\hspace{.5cm} Let $p\neq q$ be odd primes. In this section we shall consider the group 
$G=\langle g \;|\; g^{pq}=1\rangle$, denote $a=g^q$, $b=g^p$ and write $G=C_p\times C_q$,
where $C_p = \langle a\rangle$ and $C_q=\langle  b\rangle$. Theorem~\ref{teorema} above, in this contexts gives the following.

\begin{theo}\label{theo1}
Let $\,G\,=\,\left<a\right>\times \left<b\right>$ be as above and assume that $p$ and $q$ satisfy~(\ref{hypopq}). Then
the  primitive idempotents of $\F G$ are
$e_0=\widehat{G}, \ \ e_1=\hat{a}(1-\hat{b}), \ \ e_2\,=\,(1-\hat{a})\hat{b},
\ \ e_3\,=\,uv+u^2v^2\, \mbox{ and } \,e_4\,=\,uv^2+u^2v,$
where $u$ and $v$ are as in (\ref{uf2p}) and (\ref{vf2p}) above.
\end{theo}

\

We introduce some notation. For each $0\leq i\leq pq-1$, we write
\begin{equation}\label{notacao}
 g^i=g^{(i_1,i_2)}, \mbox{ where } i_1\equiv i (\!\!\!\!\!\mod p) \mbox{ and }i_2\equiv i (\!\!\!\!\!\mod q),
\end{equation}
and, for given sets $X\subset \zzz_p$ and $Y\subset \zzz_q$, we  write $g^j\in (X,Y)$ to indicate $g^j=g^{(j_1,j_2)}$, with $(j_1,j_2)\in (X,Y)$.

For a prime $p>0$, we  write $Q^p$ for the set of quadratic residues modulo $p$,
$N^p$ for the set of non quadratic residues modulo $p$ and specialize the notation~(\ref{partitionZp6sets}) to
$$Q_Q^p=\{x\in \zzz_p : x \in Q^p \;\ \mbox{and}\;\ x-1\in Q^p\},$$
$$Q_N^p  = \{x\in \zzz_p : x \in Q^p \;\ \mbox{and}\;\ x-1\in N^p\},$$
$$N_Q^p = \{x\in \zzz_p : x \in N^p \;\ \mbox{and}\;\ x-1\in Q^p\},$$  
$$N_N^p =  \{x\in \zzz_p : x \in N^p \;\ \mbox{and}\;\ x-1\in N^p\}.$$

\begin{lemma} \label{cyclasses}
Let $G$ be an abelian group of order $pq$ as in Theorem~\ref{theo1}. Then the $2$-cyclotomic classes of $G$ are:
\begin{eqnarray*}\label{4cyclotG}
\mathcal{C}_1 & = & \{1\} \\
 \mathcal{C}_{g} & = & \{g, g^2, g^{2^2},\ldots,g^{2^{\frac{(p-1)(q-1)}{2}-1}} \} \\ & = & \{g^{(i,j)} \;|\; (i,j)\in  (Q^p,Q^q)\cup (N^p,N^q)\} \\
\mathcal{C}_a & = & \mathcal{C}_{g^q}=\{g^q,g^{2q},g^{2^2q},g^{2^3q},\ldots, g^{2^{p-2}q}\} \\ 
& = & \{g^{(i,j)} \;|\; (i,j)\in  (Q^p,0)\cup (N^p,0)\} \\
 \mathcal{C}_b & = & \mathcal{C}_{g^p}=\{g^p,g^{2p},g^{2^2p},g^{2^3p},\ldots, g^{2^{q-2}p}\} \\ & = &\{g^{(i,j)} \;|\; (i,j)\in  (0,Q^q)\cup (0,N^q)\}
\end{eqnarray*}

\noindent and either
\begin{eqnarray*}
 \hspace{-0.2cm}(a) \  \mathcal{C}_{g^{p+q}}& =& \{g^{p+q},g^{2(p+q)}, 
 \ldots, g^{(2^{\frac{(p-1)(q-1)}{2}-1})(p+q)}\} \\
 & = & \{g^{(i,j)} \;|\; (i,j)\in  (Q^p,N^q)\cup (N^p,Q^q)\},
 \end{eqnarray*}
 if $p\equiv 3 \ (\!\!\!\!\mod 4)$ and  $q\equiv 3 \ (\!\!\!\!\mod 4)$ or 
%
\begin{eqnarray*} (b) \  \mathcal{C}_{g^{-1}} & = & \{g^{-1}, g^{-2}, g^{-2^2},\ldots,g^{-(2^{\frac{(p-1)(q-1)}{2}-1})} \} \hspace*{2cm} \\
& = & \{g^{(i,j)} \;|\; (i,j)\in  (Q^p,N^q)\cup (N^p,Q^q)\},
\end{eqnarray*}
if  $p\equiv 1 \ (\!\!\!\!\mod 4)$ and $q\equiv 3 \ (\!\!\!\!\mod 4)$.
\end{lemma}

\begin{IEEEproof}
Since $\bar{2}$ generates $U(\zzz_p)$ and $U(\zzz_q)$,
the least positive integer $i$ such that
$(g^q)^{2^i}=g^q$ is $p-1$ and, similarly, the least positive integer $j$ such that
$(g^p)^{2^j}=g^p$ is $q-1$. Hence $|\mathcal{C}_a|=p-1$ and $|\mathcal{C}_b|=q-1$. By Lemma~\ref{order2modpq}, we have
$|\mathcal{C}_g| = \frac{(p-1)(q-1)}{2}$.

Note that, using the notation in~(\ref{notacao}), we can write
{\small
$$\mathcal{C}_g=\left\{g^{(i,j)} \,|\, i\equiv 2^k \!\!\!\!\!\mod p, j\equiv 2^k \!\!\!\!\!\mod q, 0\leq k \leq  |\mathcal{C}_g|-1\right\}.$$ 
}
Since
 $(2,2)\in (N^p,N^q)$, all elements in $\mathcal{C}_g$ are of the form $g^{2^i}=g^{(i_1,i_2)}$, with ${(i_1,i_2)}\in (Q^p,Q^q)\cup (N^p,N^q)$. We claim that all pairs in $(Q^p,Q^q)\cup (N^p,N^q)$ do appear
as powers of elements in $\mathcal{C}_g$. This is so because
 $|(Q^p,Q^q)\cup (N^p,N^q)| = \frac{(p-1)(q-1)}{4}+ \frac{(p-1)(q-1)}{4} =\frac{(p-1)(q-1)}{2} =|\mathcal{C}_g|.$

Note that $\mathcal{C}_{g^q}\subset \{ g^{(2^i q,0)} \;|\; i\in \N\}$.  As $\bar{2}$ generates $U(\zzz_p)$ and $q$ is invertible
$\!\!\!\!\mod p$, we can write  $$\mathcal{C}_{g^q}=\{ g^{(2^i q,0)} \;|\;(2^i q,0)\in (Q^p,0)\cup  (N^p,0)\}.$$
Similarly,
 $\mathcal{C}_{g^p}=\{ g^{(0,2^i p)} : (0,2^i p)\in (0,Q^q)\cup  (0,N^q)\}$.\\

To determine the last cyclotomic class, we shall consider two separate cases.\\
{\em Case 1. $p\equiv 3 \ (\!\!\!\!\mod 4)$ and $q\equiv 3 \ (\!\!\!\!\mod 4)$.} By the  Quadratic Reciprocity Law
we have  $p\in Q^q \ (\!\!\!\!\mod q)$ if and only if
$q\in N^p \ (\!\!\!\!\mod p)$, hence the element $g^{p+q}=g^{(q,p)}\in (N^p,Q^q)\cup (Q^p,N^q)$ so $g^{p+q}\not\in \mathcal{C}_g$.
Obviously, $1\neq g^{p+q}\not\in \mathcal{C}_{g^p}\cup \mathcal{C}_{g^q}$. Consequently, in this case, the fifth $2$-cyclotomic class is:
$$ \mathcal{C}_{g^{p+q}}=\left\{g^{p+q},g^{2(p+q)},g^{2^2(p+q)},\ldots, g^{(2^{\frac{(p-1)(q-1)}{2}-1})(p+q)}\right\}. $$
{\em Case 2. $p\equiv 1 \ (\!\!\!\!\mod 4)$ and $q\equiv 3 \ (\!\!\!\!\mod 4)$.}
By Lemma~\ref{-1rq}, we have $-1\in Q^p$ and $-1\in N^q$.
Thus $g^{-1}=g^{(-1,-1)}\in  (Q^p,N^q)$ so $ g^{-1}\not\in \mathcal{C}_g$. By considering the order of the corresponding elements, we also have $1\neq g^{-1}\not\in \mathcal{C}_{g^p}\cup \mathcal{C}_{g^q}$.

Therefore we can describe the fifth $2$-cyclotomic class as:

$$\,\mathcal{C}_{g^{-1}}=\left\{g^{-1}, g^{-2}, g^{-2^2},\ldots,g^{-(2^{\frac{(p-1)(q-1)}{2}-1})} \right\}.$$
\end{IEEEproof}

For  an element $x\in G$, we shall write $S_x=\displaystyle \sum_{h\in \mathcal{C}_x} h$,  to denote the sum of all elements in the $2$-cyclotomic class of $x$ in $G$. Then we can write:

In Case 1 above:
\begin{equation}\label{e3e4caso1}
e_3 = {S_g}+{S_{g^p}}+{S_{g^q}} \quad\mbox{ and }\quad e_4  = {S_{g^{p+q}}}+{S_{g^p}}+{S_{g^q}}.
\end{equation}
and, in Case 2:
\begin{equation}\label{e3e4caso2}
e_3  =  {S_{g}}+{S_{g^q}}  \quad\mbox{ and }\quad e_4  =  {S_{g^{-1}}}+{S_{g^q}}.
\end{equation}


\begin{prop}\label{bases} With the same hypothesis as in Theorem~\ref{theo1} we have:
\begin{enumerate}
\item[$(i)$]  $\{e_0\}$ is a basis of $(\F_2 G)e_0.$
\item[$(ii)$] 
${\mathcal B}_1= \{ \hat{a}(b^j - 1) \;|\; 1\leq j \leq q-1\} \mbox{ and }
{\mathcal B}_1'=\{ b^je_1 \;|\; 1\leq j \leq q-1\}$
are bases of $(\F_2 G)e_1.$

\item[$(iii)$] 
${\mathcal B}_2 =\{ (a^j - 1)\hat{b} \;|\; 1\leq j \leq p-1\}
\mbox{ and }
{\mathcal B}_2' = \{ a^je_2 \;|\; 1\leq j \leq p-1\}$
are bases of $(\F_2 G)e_2.$
\end{enumerate}
Let  $s, t\in \zzz$ be such that $sq\equiv 1 (\!\!\!\!\mod p)$
and $tp\equiv 1 (\!\!\!\!\mod q)$, then:

\begin{enumerate}
\item[$(iv)$]  For $ y=(1+a^s)(1+b^t)e_3\in (\F_2 G)
e_3$, the set $\{y, gy, g^2y,\ldots, g^{\frac{(p-1)(q-1)}{2}-1}y\}$ is a basis of $(\F_2 G)
e_3$.

\item[$(v)$]  For $ y=(1+a^s)(1+b^t)e_4\in (\F_2 G)
e_4$, the set $\{y, gy, g^2y,\ldots, g^{\frac{(p-1)(q-1)}{2}-1}y\}$  is a basis of $(\F_2 G)
e_4$.
\end{enumerate}
\end{prop}

\begin{IEEEproof}
The validity of $(i)$ is obvious. To prove $(ii)$, notice that
$$(\F_2G)e_1  =  (\F_2G)\hat{a}(1-\hat{b})  \cong  (\F_2C_q)(1-\hat{b}) $$
 and this isomorphism maps the element $x\in (\F_2C_q)(1-\hat{b})$ to $x\hat{a}\in (\F_2G)e_1$. As the set $\{ b^j -1 \;|\; 0<j\leq q-1\}$ is a basis of $(\F_2C_q)(1-\hat{b})$ (see \cite[Proposition 3.2.10, p.133]{MS}), it follows that ${\mathcal B}_1'$ is a basis of $(\F_2 G) e_1.$

 To prove that ${\mathcal B}_1$ is also a basis of $(\F_2 G) e_1$, we  prove first that $\{b^j-\hat{b} \;|\; 1\leq j \leq q-1\}$ is  a basis of $(\F_2C_q)(1-\hat{b}) $. To do so, it suffices to show that it is linearly independent, as it contains precisely $q-1$ elements.

 Assume that there exist coefficients $x_j\in \F_2$, $1\leq j\leq q-1$, such that $\sum_{j=1}^{q-1} x_j(b^j-\hat{b}) = 0$. If $\sum_{j=1}^{q-1} x_j=0$, then $\sum_{j=1}^{q-1} x_jb^j =0$ so $x_j=0$, for all $j$, $1\leq j\leq q-1$. If $\sum_{j=1}^{q-1} x_j=1$, then $\sum_{j=1}^{q-1} x_jb^j + \hat{b}=0$ so we must have $x_j = 1$, for all $1\leq j\leq q-1$, which implies  $\sum_{j=1}^{q-1} x_j=0$, a contradiction.

 Because of the isomorphism above, it follows that also ${\mathcal B}_1$ is a basis of $(\F_2G)e_1$.

 The proof of $(iii)$ is similar.

 To prove $(iv)$, notice that, by Lemma~(\ref{lemars}), $\dim_{\F_2} [(\F_2G)e_3] = \frac{(p-1)(q-1)}{2}$. Also,   
 $(\F_2G)e_3 = \F_2 (ge_3)$ is a finite field and $ge_3$  a root of an irreducible polynomial of degree $\frac{(p-1)(q-1)}{2}$. Hence,
 the set $\{ e_3, ge_3, g^2e_3, \ldots, g^{\frac{(p-1)(q-1)}{2}-1}e_3\}$ is a basis of $(\F_2G)e_3$.

 We shall prove independently in Lemma~\ref{pesodee3} that the element $$y=(1+a^s)(1+b^t)e_3 \in (\F_2G)e_3$$ is nonzero. Then
 $\{ y, gy, g^2y, \ldots, g^{\frac{(p-1)(q-1)}{2}-1}y\}$ is also a basis of $(\F_2G)e_3$.

 The proof of $(v)$ is a consequence of the isomorphism $(\F_2 G)e_3\cong (\F_2 G)e_4$.
\end{IEEEproof}

\begin{cor}\label{dim} Let $G$ be  as above. The dimensions of the minimal ideals of $\F_2 G$ are:
\begin{itemize}
\item[$(i)$] $\displaystyle\dim_{\F_2}[(\F_2 G)e_0]\,=\,1.$

\item[$(ii)$]  $\,\displaystyle\dim_{\F_2}[(\F_2 G)e_1]\,=\,q-1.$

\item[$(iii)$]  $\,\displaystyle\dim_{\F_2}[(\F_2 G)e_2]\,=\,p-1.$

\item[$(iv)$]  $\,\displaystyle\dim_{\F_2}[(\F_2 G)e_3]\,=\,(p-1)(q-1)/2.$

\item[$(v)$] $\,\displaystyle\dim_{\F_2}[(\F_2 G)e_4]\,=\,(p-1)(q-1)/2.$
\end{itemize}
\end{cor}

We now compute the weight of a particular element of $\F_2G$.

\begin{lemma}\label{pesodee3}
With the same hypothesis of the Theorem~\ref{theo1} and notation above, the element $$ y=(1+g^{sq})(1+g^{tp})e_3=(1+a^s)(1+b^t)e_3\in(\F_2G)e_3,$$ with
$s, t\in \zzz$ such that $sq\equiv 1 (\!\!\!\!\mod p)$ and
$tp\equiv 1 (\!\!\!\!\mod q)$, has weight $p+q$.
\end{lemma}
\begin{IEEEproof}
Since $p\in U(\zzz_q)$ and $q\in U(\zzz_p)$, by~\cite[Theorem 69]{Landau},
we can choose $s, t\in \zzz^*$ such that $sq\equiv 1$ $(\!\!\!\!\mod p)$
and  $tp\equiv 1 (\!\!\!\!\mod q)$. For this choice of $s,t\in \zzz^*$, we have $ y=(1+g^{sq})(1+g^{tp})e_3=(1+g^{(1,0)})(1+g^{(0,1)})e_3 \in (\F_2G) e_3$.

We shall consider two cases.

\textbf{Case 1:} $p\equiv 3 (\!\!\!\!\mod 4)$ and $q\equiv 3 (\!\!\!\!\mod 4)$. In this case, by~(\ref{e3e4caso1}), we have 
\begin{eqnarray*} e_3 & = & \underbrace
{(g+ g^2+ g^{2^2}+\cdots +g^{2^{\frac{(p-1)(q-1)}{2}-1}})}_{S_{g}} \\ & + &\underbrace{(g^p+ g^{2p}+\cdots+ g^{(q-1)p})}_{S_{g^p}}+ \\
& & +\underbrace{(g^q+ g^{2q}+\cdots+ g^{(p-1)q})}_{S_{g^q}}.
\end{eqnarray*}
As in the proof of Lemma~\ref{cyclasses}, $S_{g}$ is the sum of all elements of $G$ with exponents in
$(Q^p,Q^q)\cup (N^p,N^q)$, $S_{g^p}$ is the sum of all elements with exponents in $(0,Q^q)\cup  (0,N^q)$ and $S_{g^q}$ is the sum of all elements with exponents in $ (Q^p,0)\cup  (N^p,0)$. Thus $\omega(e_3)=\frac{(p-1)(q-1)}{2}+(p-1)+(q-1)$.

Now we set
$e_3=S_{g}+(1+S_{g^p})+(1+S_{g^q})$  and 
claim that $y[(1+S_{g^p})+(1+S_{g^q})]=0$. Indeed,\\ 
$(1+g^{(1,0)})(1+g^{(0,1)})[(1+S_{g^p})+(1+S_{g^q})]  =$ \\
$  (1+S_{g^p})+(1+S_{g^q})+ g^{(1,0)}[(1+S_{g^p})+(1+S_{g^q})] +  $ \\
$   + g^{(0,1)}[(1+S_{g^p})+(1+S_{g^q})]+g^{(1,1)}[(1+S_{g^p})+(1+S_{g^q})]  $\\
$= (1+S_{g^p})+(1+S_{g^q}) +  g^{(1,0)}(1+S_{g^p})+(1+S_{g^q}) + $  \\
$(1+S_{g^p})+g^{(0,1)}(1+S_{g^q})  +  g^{(1,1)}(1+S_{g^p})+g^{(1,1)}(1+S_{g^q}) $    \\
$= g^{(1,0)}(1+S_{g^p})+g^{(0,1)}(1+S_{g^q}) +  g^{(1,1)}(1+S_{g^p})+g^{(1,1)}(1+S_{g^q})=0.  $

\

To prove the last equality it is convenient to write all the elements in the form $g^{(i_1,i_2)}$.
In this way it is clear that each element appears twice in the sum, hence its coefficient is zero.

Thus $\omega(y)=\omega(yS_g)$ and it is enough to compute
the weight of $yS_g$. To do so, note that $\zzz_p$  is the union of six disjoint subsets as in Remark~\ref{6conjuntos}
and, for $\zzz_p^*\times \zzz_q^*$, we have:
\vspace{0.2cm}

\noindent $ N^p\times N^q\,=\,\{(N_Q^p,N_Q^q),(N_Q^p,N_N^q),(N_N^p,N_Q^q),(N_N^p,N_N^q)\}$ \\
\noindent and \\
\noindent $Q^p\times Q^q=\{(Q_Q^p,Q_Q^q),(Q_Q^p,Q_N^q),(Q_Q^p,1),(Q_N^p,Q_Q^q),$
$  \hspace{2.2cm}(Q_N^p,Q_N^q),(Q_N^p,1),(1,Q_Q^q),(1,Q_N^q),(1,1)\}.$

\vspace{0.2cm}

Now we get

\vspace{0.2cm}

{\small
\noindent $(1,0) + (Q^p\times Q^q)=\{(N_Q^p,Q_Q^q),(N_Q^p,1), (N_Q^p,Q_N^q),(Q_Q^p,Q_Q^q),$ \\
$\hspace{3.1cm} (Q_Q^p,1), (Q_Q^p,Q_N^q)\}, $

\

\noindent $(1,0) +(N^p\times N^q)=\{(Q_N^p,N_Q^q),(Q_N^p,N_N^q),(0,N_Q^q),(0,N_N^q),  $ \\
\noindent $\hspace{3.1cm} (N_N^p,N_Q^q),(N_N^p,N_N^q)\},$

\

\noindent $(0,1) + ((Q^p\times Q^q)\cup (N^p\times N^q))=\{(Q_Q^p,N_Q^q),(1,N_Q^q)\}\cup $ 

$\hspace{0.8cm}\cup\{ (Q_N^p,N_Q^q),(Q_Q^p,Q_Q^q),(1,Q_Q^q),(Q_N^p,Q_Q^q),(N_Q^p,Q_N^q)\}$

$\hspace{0.8cm} \cup \{(N_N^p,Q_N^q),(N_Q^p,0),(N_N^p,0),(N_Q^p,N_N^q),(N_N^p,N_N^q)\},$

\

\noindent $(1,1) + ((Q^p\times Q^q)\cup (N^p\times N^q))=\{(0,0),(Q_Q^p,Q_Q^q),(Q_Q^p,N_Q^q)\}$ 

$\hspace{0.8cm} \cup \{(N_Q^p,Q_Q^q),(N_Q^p,N_Q^q),(Q_N^p,Q_N^q),(Q_N^p,N_N^q),(Q_N^p,0)\}$ 

$\hspace{0.8cm} \cup \{(N_N^p,Q_N^q),(N_N^p,N_N^q),(N_N^p,0),(0,N_N^q),(0,Q_N^q)\}$
}

\

Thus the elements that appear in $yS_g$ are all those with exponents in the set
$\{(Q_N^p,1),(1,Q_N^q), (N_Q^p,1), (0,N_Q^q),$ $(1,N_Q^q),(N_Q^p,0),(Q_N^p,0),(0,Q_N^q),(1,1),(0,0)\},$
whose cardinality, by Lemma~\ref{sizesets}, is $2+2|Q_N^p|+2|Q_N^q|+2|N_Q^p|+2|N_Q^q|=2+ 2\frac{p+1}{4}+2\frac{q+1}{4}+2\frac{p-3}{4}+2\frac{q-3}{4}=p+q$.
\vspace{0.3cm}

\textbf{Case 2:} $p\equiv 1 (\!\mod 4)$ and $q\equiv 3 (\!\mod 4)$.
In this case, by~(\ref{e3e4caso2}), we have 

\begin{eqnarray*}
e_3 & = & \underbrace
{(g+ g^2+ g^{2^2}+\cdots +g^{2^{\frac{(p-1)(q-1)}{2}}-1})}_{S_{g}} \\ & + & \underbrace{(g^q+ g^{2q}+\cdots+ g^{(p-1)q})}_{S_{g^q}}
\end{eqnarray*}
where, as in the proof of Lemma~\ref{cyclasses}, $S_{g}$ is the sum of all elements of $G$ with exponents in
$(Q^p,Q^q)\cup (N^p,N^q)$ and $S_{g^q}$ is the sum of all element with exponents in $ (Q^p,0)\cup  (N^p,0)$. Thus $\omega(e_3)=\frac{(p-1)(q-1)}{2}+(p-1)$.

We claim that $\omega(y)=p+q$. Indeed, we can rewrite $S_{g^q}$ as  $S_{g^q}= g^{(1,0)}+ g^{(2,0)}+\cdots + g^{(p-1,0)}$.
Thus \\
$(1+g^{(1,0)})(1+g^{(0,1)})S_{g^q}=g^{(1,0)}+ g^{(2,0)}+\cdots + g^{(p-1,0)}+ g^{(2,0)}+ g^{(3,0)}+\cdots + g^{(0,0)}+g^{(1,1)}+ g^{(2,1)}+\cdots + g^{(p-1,1)}+g^{(2,1)}+ g^{(3,1)}+\cdots + g^{(0,1)}=1+g^{(1,0)}+g^{(0,1)}+g^{(1,1)}$.

In $(1+g^{(1,0)})(1+g^{(0,1)})S_{g}$, as shown in Case 1, all the elements have exponents in the set
$\{(Q_N^p,1),(1,Q_N^q), (N_Q^p,1),$ $ (0,N_Q^q), (1,N_Q^q),(N_Q^p,0),(Q_N^p,0),(0,Q_N^q),(1,1),(0,0)\}$.
Therefore,  the elements that appear in $y$
are all those with exponents in the set
$\{(Q_N^p,1),(1,Q_N^q),(1,0), (N_Q^p,1), (0,N_Q^q),$ $(1,N_Q^q),(N_Q^p,0),(Q_N^p,0),(0,Q_N^q),(0,1)\}$,
whose cardinality, by Lemma~\ref{sizesets},
is also $2+2|Q_N^p|+2|Q_N^q|+2|N_Q^p|+2|N_Q^q|=2+ 2\frac{p-1}{4}+2\frac{q-3}{4}+2\frac{p-1}{4}+2\frac{q+1}{4}=p+q$.

\end{IEEEproof}

\begin{remark}
 Lemma~\ref{pesodee3}, shows that the elements in the bases defined in parts $(iv)$ and $(v)$ of Proposition~\ref{bases}   have all the same weight $p+q$.
\end{remark}

\begin{theo} \label{dimweightcpcq} Let $G\,=\,< g >$ be an abelian group of order $pq$ as in Theorem~\ref{theo1}.
Then:

\noindent $(i)$  $\,\omega((\F_2 G) e_0)\,=\,p\,q$.

\noindent $(ii)$ $\,\omega((\F_2 G) e_1)\,=\,2\,p$.

\noindent $(iii)$ $\,\omega((\F_2 G) e_2)\,=\,2\,q$.

\noindent $(iv)$ $\,4\leq \omega((\F_2 G) e_3)\,\leq p+q$.

\noindent $(v)$ $\,4\leq \omega((\F_2 G) e_4)\,\leq p+q$.
\end{theo}
\begin{IEEEproof} $(i)$ follows immediately as $(\F_2 G)
e_0\cong \F_2$.

$(ii)$ Recall that  $e_1= \hat{a}(1-\hat{b})$.
Since $\,(b+b^2)e_1\,=\,(b+b^2)\hat{a}\in (\F_2 G)e_1$ and $supp(b\hat{a})\cap supp(b^2\hat{a})=\emptyset$ then $\,\omega((b+b^2)e_1)=2p$. Hence $\,\omega((\F_2 G) e_1)\leq 2p$.

An arbitrary element $\alpha\in (\F_2G)e_1$ is of the form
$$\alpha=\sum_{i,j}k_{ij}a^ib^j\hat{a}(1-\hat{b})= \left[\sum_{j=0}^{q-1}(\sum_{i=0}^{p-1}k_{ij})b^j(1-\hat{b})\right]\hat{a},
 $$
with $k_{i,j}\in \F_2$, hence is also an element of $(\F_2 G)\hat{a}$. 
An element $\beta\in(\F_2 G)\hat{a}$ is of the form
$$\beta=\sum_{i,j}\ell_{ij}a^ib^j\hat{a}=\sum_{j=0}^{q-1}(\sum_{i=0}^{p-1}\ell_{ij})b^j\hat{a}, $$  
with $\ell_{i,j}\in \F_2$.
Thus a nonzero element $\beta\in(\F_2 G)\hat{a}$ has weight $\omega(\beta)=np$, with $n\geq 1$, as the
elements $b^j\hat{a}$, for different values of $j$ have disjoint supports.

Now as $b^j\hat{a} e_1=b^je_1\neq b^j\hat{a}$, the element $b^j\hat{a}\not\in (\F_2 G) e_1$.
Hence, for an
element $\alpha \in (\F_2 G) e_1$ to have weight $\,p$, we must have
$\,\alpha=b^j\hat{a}$, for some $j$, a contradiction.
Therefore, $\,2p\,$ is the minimum weight of the code $(\F_2 G) e_1$.

  $(iii)$ follows as $(ii)$  interchanging
$\,a\,$ with $\,b\,$ and $\,p\,$ with
$\,q$.

For $(iv)$ and $(v)$ it is enough to compute the weight of one of these codes, since there exists an automorphism
of $\F_2 G$ induced by a group automorphism of $G$ that maps one code into the other, hence they are equivalent.

As $(1+a)(1+b)(e_3+e_4)=(1+a)(1+b)(1+\hat{a})(1+\hat{b})=(1+a)(1+b)$, then $(1+a)(1+b)\in (\F_2 G)(e_3+e_4)$. Besides,
it is easy to prove that there is no element of weight $2$ in $(\F_2 G)(e_3+e_4)$ and,
as $e_3, e_4\in (\F_2 G)(e_3+e_4)$, then $4\leq \omega[(\F_2 G)(e_3+e_4)]\leq \omega[(\F_2 G)e_j]$, for $j=3,4$.
By Lemma~\ref{pesodee3}, we have $\omega[(\F_2 G)e_3]\leq p+q$.

\end{IEEEproof}

%

\subsection{Examples}

\begin{example}
The upper bound for the weights of the codes in parts (iv) and (v) of Theorem~\ref{dimweightcpcq}  is  sharp,  as it is attained by the code generated by the primitive idempotent 
$$e= g+g^2+g^3+g^4+g^6+g^8+g^9+g^{12}\in \F_2 C_{15}.$$ Indeed, the
group code $I=(\F_2 C_{15})e$
 has dimension $4$ over $\F_2$ and it is easy to see that
$I\,=\,\{g^j\ e\,|\,j=0,\ldots, 14\}\cup\{0\}$. 
Hence all non zero elements in $I$ have  weight equal to $\omega(e)= 8$. 
\end{example}

However, this is not always the case as we can
see  below.

\begin{example}
Let $C_{33}=\langle g\;|\; g^{33}=1 \rangle$ be the cyclic group with $33$ elements and  $(\F_2 C_{33})e$ be the group code generated by the primitive idempotent
$e=g+g^2+g^3+g^4+g^6+g^8+g^9+g^{11}+g^{12}+g^{15}+g^{16}+g^{17}+g^{18}+g^{21}+g^{22}+g^{24}+g^{25}+g^{27}+g^{29}+g^{30}+g^{31}+g^{32}$. Then the
weight distribution of $(\F_2 C_{33})e_3$
is as follows:
$$
\begin{array} {|c|c|c|c|c|c|c|}
\hline
\mbox{Vector Weight} & 12 & 14 & 16 & 18 & 20 & 22  \\ \hline
\mbox{Number of Vectors} & 165 & 165 & 165 & 330 & 165 & 33 \\ \hline
\end{array}
$$
\end{example}

In fact, notice first  the ideal $(\F_2 C_{33})e$ is a field and Corollary~\ref{dim} shows that its dimension over $\F_2$ is $10$, so
its group of units, $U((\F_2 C_{33})e)$, has order $\F_2^{10}-1=1023= 33\cdot 31$.

Notice that $C_{33} \cong C_{33}\cdot e \subset  U((\F_2 C_{33})e)$. Also $$((g+g^{-1})e)^{32} = (g^{-1} +g )e,$$ 
thus $x=(g+g^{-1})e$ is an element of order equal to either $1$ or $31$ inside $U((\F_2 C_{33})e)$. But $x\neq e$, as $\omega(x)=18$ and $\omega(e)=22$. Hence $U((\F_2 C_{33})e) = C_{33}\cdot e \times \langle x\rangle$.

Computing the $2$-cyclotomic classes in $\langle x\rangle$ we get:

$U^*_0=\{0\}$,

$U^*_1=\{x,x^2,x^4,x^8,x^{16}\}$, 

$U^*_2=\{x^3,x^6,x^{12},x^{24},x^{17}\}$,  

 $U^*_3=\{x^5,x^{10},x^{20},x^{9},x^{18}\}$,
 
$U^*_4=\{x^7,x^{14},x^{28},x^{25},x^{19}\}$, 

 $U^*_5=\{x^{11},x^{22},x^{13},x^{26},x^{21}\}$ and 
 
$U^*_6=\{x^{15},x^{30},x^{29},x^{27},x^{23}\}$.

For a fixed $0\leq k\leq 31$, we have $\omega(g^jx^k)=\omega(x^k)$, for all $0\leq j\leq 32$ and
for each $0\leq t\leq 6$, all $y\in U^*_t$ have the same weight.

Using these facts to compute the weights, we have that:
\begin{itemize}
\item There are $33$ distinct elements of weight $22$ in $(\F_2 C_{33})e_3$, since $\omega(e_3)=22$.
\item There are $330$ distinct elements in $(\F_2 C_{33})e_3$ with weight $18$, as $\omega(x)=18$, $\omega(x^5)=18$ and
$U^*_1\cap U^*_3=\emptyset$.
\item There are $165$ distinct elements in $(\F_2 C_{33})e_3$ with weight $16$, since $\omega(x^3)=16$.
\item There are $165$ distinct elements in $(\F_2 C_{33})e_3$ with weight $20$, since $\omega(x^7)=20$.
\item There are $165$ distinct elements in $(\F_2 C_{33})e_3$ with weight $12$, since $\omega(x^{11})=12$.
\item There are $165$ distinct elements in $(\F_2 C_{33})e_3$ with weight $14$, since $\omega(x^{15})=14$.
\end{itemize}

\begin{theo}\label{3primes}
Let $\,p_1,\,p_2\,$ and $\,p_3\,$ be three distinct positive odd prime numbers such that
$\,\gcd (p_i-1,\,p_j-1)=2$, for $1\leq i\neq j\leq 3$, and $\,\bar{2}\;
\mbox{generates the groups of units}\; U(\zzz_{p_i})$. Then the primitive idempotents of
the group algebra $\F_2G$ \linebreak for the finite abelian group $\,G=C_{p_1}\times C_{p_2}\times C_{p_3}$, with $C_{p_1}=<a>$, $C_{p_2}=<b>$ and
$C_{p_2}=<c>$, are

$e_0=\hat{a}\hat{b}\hat{c}$, 
$e_1=\hat{a}\hat{b}(1-\hat{c})$, 
$e_2=\hat{a}(1-\hat{b})\hat{c}$,
$e_3=(1-\hat{a})\hat{b}\hat{c}$,

$e_4=(uv+u^2v^2)\hat{c}$,
$e_5=(u^2v+uv^2)\hat{c}$

$e_6=(uw+u^2w^2)\hat{b}$, 
$e_7=(u^2w+uw^2)\hat{b}$

$e_8=(vw+v^2w^2)\hat{a}$,
$e_9=(v^2w+vw^2)\hat{a}$

$e_{10}=(1-\hat{a})(1-\hat{b})(1-\hat{c})+ u^2v^2w+uvw^2$

$e_{11}=(1-\hat{a})(1-\hat{b})(1-\hat{c})+ u^2v^2w^2+uvw$

$e_{12}=(1-\hat{a})(1-\hat{b})(1-\hat{c})+ u^2vw^2+uv^2w^2$

and

$e_{13}=(1-\hat{a})(1-\hat{b})(1-\hat{c})+ uv^2w+u^2vw^2$,

\noindent where $u,v$ are defined as in (\ref{uf2p}) and (\ref{vf2p}), respectively, and

\begin{equation}\label{wf2p3}
w\,=\,\left\{
\begin{array}{ll}
c^{2^0}+c^{2^2}+\cdots + c^{2^{p-3}}, &
\mbox{if } p_3\equiv 1 (\!\!\!\!\!\mod 4)\;\mbox{or}\\
1+c^{2^0}+c^{2^2}+\cdots  +c^{2^{p-3}},
& \mbox{if } p_3\equiv 3 (\!\!\!\!\!\mod 4) \end{array}\right.
\end{equation}
\end{theo}
\begin{IEEEproof}
As in the proof of Theorem~\ref{theo1}, by~\cite[Lemma 3.1]{FM}, for each $i=1,2,3$, $\F_2 C_{p_i}$ contains two primitive idempotents, namely, $\wh{C_{p_i}}$ and $1-\wh{C_{p_i}}$ so
$$\F_2C_{p_i}  \cong  (\F_2 C_{p_i})\cdot \widehat{C_{p_i}} \,\oplus
(\F_2 C_{p_i})\cdot (1-\widehat{C_{p_i}}) \label{f2cpi} \cong \F_2 \oplus \F_{2^{p_i-1}}.$$
By Lemma~\ref{lemars} we have $$\F_{2^{\frac{(p_1-1)(p_2-1)}{2}}}\otimes \F_{2^{p_3-1}}\cong 2\cdot \F_{2^{\frac{(p_1-1)(p_2-1)(p_3-1)}{4}}},$$
 thus:
\begin{eqnarray}\label{simple3primos} 
\F_2G & \cong & \F_2[C_{p_1}\times C_{p_2}\times C_{p_3}]\nonumber \\ & \cong &
 (\F_2C_{p_1} \otimes_{\F_2}\F_2C_{p_2}) \otimes_{\F_2}\F_2C_{p_3} \nonumber\\
& \cong & (\F_2\oplus \F_{2^{p_1-1}}\,\oplus
\,\F_{2^{p_2-1}} \,\oplus 2\cdot
\F_{2^{\frac{(p_1-1)(p_2-1)}{2}}}) \nonumber \\ && \otimes_{\F_2} (\F_2\oplus \F_{2^{p_3-1}} )\\
& \cong & \F_2\oplus \F_{2^{p_1-1}}\,\oplus
\,\F_{2^{p_2-1}} \,\oplus 2\cdot
\F_{2^{\frac{(p_1-1)(p_2-1)}{2}}}   \nonumber\\
& & \oplus \F_{2^{p_3-1}} \oplus 2\cdot \F_{2^{\frac{(p_1-1)(p_3-1)}{2}}}\oplus 
2\cdot \F_{2^{\frac{(p_2-1)(p_3-1)}{2}}} \nonumber \\ && \oplus 2\cdot (\F_{2^{\frac{(p_1-1)(p_2-1)}{2}}}\otimes \F_{2^{p_3-1}}).\nonumber
\end{eqnarray}
Therefore, there exist $14$ simple components in this decomposition.  First note that
\begin{eqnarray*} & & \hat{a}\ \hat{b}\ \hat{c} \  +  \ (1-\hat{a}) \ \hat{b} \ \hat{c} \  + \ \hat{a} \ (1-\hat{b}) \ \hat{c}\  + \ \hat{a} \ \hat{b}\ (1-\hat{c}) \
 + \\
 && + \ (1-\hat{a}) \ (1-\hat{b}) \ \hat{c}  + \ \hat{a} \ (1-\hat{b})\ (1-\hat{c})\\ && + \ (1-\hat{a})\ \hat{b} \ (1-\hat{c})\  +\  (1-\hat{a})\ (1-\hat{b})\ (1-\hat{c})\ =\ 1.
\end{eqnarray*}

These components are  as follows:
\begin{eqnarray*}
\F_2G\cdot\hat{a}\hat{b}\hat{c} & = & \F_2G\cdot \widehat{G}\ \cong  \ \F_2 \\
\F_2G\cdot (1-\hat{a})\hat{b}\hat{c} & \cong & \F_2G\cdot (1-\widehat{C_{p_1}})\widehat{C_{p_2}}\widehat{C_{p_3}} \\ & \cong & \  \F_2 C_{p_1} (1-\widehat{C_{p_1}})\  \cong \ \F_{2^{p_1-1}} \\
\F_2G\cdot\hat{a}(1-\hat{b})\hat{c} & \cong & \F_2G\cdot \widehat{C_{p_1}}(1-\widehat{C_{p_2}})\widehat{C_{p_3}} \\ & \cong & \ \F_2 C_{p_2} (1-\widehat{C_{p_2}})  \ \cong \ \F_{2^{p_2-1}} \\
\F_2G\cdot\hat{a}\hat{b}(1-\hat{c}) & \cong & \F_2G\cdot \widehat{C_{p_1}}\widehat{C_{p_2}}(1-\widehat{C_{p_3}}) \\ & \cong & \  \F_2 C_{p_3} (1-\widehat{C_{p_3}}) \ \cong \ \F_{2^{p_3-1}}.
\end{eqnarray*}
Therefore
\begin{eqnarray}\label{simpleideals}
\F_2G & \cong & \F_2C_{p_1} \otimes_{\F_2}\F_2C_{p_2} \otimes_{\F_2}\F_2C_{p_3} \nonumber\\
& \cong & (\F_2C_{p_1}\hat{a}\oplus \F_2C_{p_1}(1-\hat{a}))\nonumber \\  &  \otimes &
(\F_2C_{p_2}\hat{b}\oplus \F_2C_{p_2}(1-\hat{b})) \nonumber \\ &  \otimes &
(\F_2C_{p_3}\hat{c}\oplus \F_2C_{p_3}(1-\hat{c})) \nonumber\\
& \cong & (\F_2C_{p_1}\hat{a}\otimes \F_2C_{p_2}\hat{b}\otimes \F_2C_{p_3}\hat{c}) \nonumber \\ &  \oplus &
(\F_2C_{p_1}\hat{a}\otimes \F_2C_{p_2}\hat{b}\otimes \F_2C_{p_3}(1-\hat{c})) \nonumber\\
&  \oplus &
(\F_2C_{p_1}\hat{a}\otimes \F_2C_{p_2}(1-\hat{b})\otimes \F_2C_{p_3}\hat{c})  \nonumber \\ & \oplus & 
(\F_2C_{p_1}\hat{a}\otimes \F_2C_{p_2}(1-\hat{b})\otimes \F_2C_{p_3}(1-\hat{c}))\nonumber\\
& \oplus &   (\F_2C_{p_1}(1-\hat{a})\otimes \F_2C_{p_2}\hat{b}\otimes \F_2C_{p_3}\hat{c})  \nonumber \\  & \oplus &
(\F_2C_{p_1}(1-\hat{a})\otimes \F_2C_{p_2}\hat{b}\otimes \F_2C_{p_3}(1-\hat{c}))\nonumber\\
& \oplus  &
(\F_2C_{p_1}(1-\hat{a})\otimes \F_2C_{p_2}(1-\hat{b})\otimes \F_2C_{p_3}\hat{c})  \nonumber\\
& \oplus &
(\F_2C_{p_1}(1-\hat{a})\otimes \F_2C_{p_2}(1-\hat{b})\otimes \F_2C_{p_3}(1-\hat{c}) )\nonumber
\end{eqnarray}
and 
\begin{eqnarray*}
\F_2 & \cong & \F_2G\cdot \widehat{G} = \F_2G\cdot\hat{a}\hat{b}\hat{c} \\ &\cong & \F_2C_{p_1}\hat{a}\otimes \F_2C_{p_2}\hat{b}\otimes \F_2C_{p_3}\hat{c}\\
\F_{2^{p_1-1}} & \cong & \F_2G\cdot (1-\widehat{C_{p_1}})\widehat{C_{p_2}}\widehat{C_{p_3}} = \F_2G\cdot (1-\hat{a})\hat{b}\hat{c}\\ & \cong & \F_2C_{p_1}(1-\hat{a})\otimes \F_2C_{p_2}\hat{b}\otimes \F_2C_{p_3}\hat{c} \\
\F_{2^{p_2-1}} & \cong & \F_2G\cdot \widehat{C_{p_1}}(1-\widehat{C_{p_2}})\widehat{C_{p_3}} =  \F_2G\cdot\hat{a}(1-\hat{b})\hat{c}\\ &  \cong &
\F_2C_{p_1}\hat{a}\otimes \F_2C_{p_2}(1-\hat{b})\otimes \F_2C_{p_3}\hat{c}  \\
\F_{2^{p_3-1}} & \cong & \F_2G\cdot \widehat{C_{p_1}}\widehat{C_{p_2}}(1-\widehat{C_{p_3}})=  \F_2G\cdot\hat{a}\hat{b}(1-\hat{c})\\ &  \cong & \F_2C_{p_1}\hat{a}\otimes \F_2C_{p_2}\hat{b}\otimes \F_2C_{p_3}(1-\hat{c})
\end{eqnarray*}
Let $ 0\neq u\in \F_2C_{p_1}(1-\hat{a})$ be an element such that $u^3=(1-\hat{a})$ and $u\neq (1-\hat{a})$;
$ 0\neq v\in \F_2C_{p_2}(1-\hat{b})$  such that $v^3=(1-\hat{b})$ and $v\neq (1-\hat{b})$ and
$ 0\neq w\in \F_2C_{p_3}(1-\hat{c})$ such that $w^3=(1-\hat{c})$ and $w\neq (1-\hat{c})$.

Then:
\begin{eqnarray*}
& & \F_2C_{p_1}(1-\hat{a})\otimes \F_2C_{p_2}(1-\hat{b})\otimes \F_2C_{p_3}\hat{c} \\ & = & (\F_2G) e_3^{ab}\hat{c}\oplus (\F_2G) e_4^{ab}\hat{c},\\
& & \F_2C_{p_1}(1-\hat{a})\otimes \F_2C_{p_2}\hat{b}\otimes \F_2C_{p_3}(1-\hat{c}) \\ & = & (\F_2G) e_3^{ac}\hat{b}\oplus (\F_2G) e_4^{ac}\hat{b},\\
& & \F_2C_{p_1}\hat{a}\otimes \F_2C_{p_2}(1-\hat{b})\otimes \F_2C_{p_3}(1-\hat{c}) \\ & = & (\F_2G) e_3^{bc}\hat{a}\oplus (\F_2G) e_4^{bc}\hat{a},
\end{eqnarray*}
where $e_3^{ab}= uv+u^2v^2$ and $e_4^{ab}= u^2v+uv^2$, $e_3^{ac}= uw+u^2w^2$ and $e_4^{ac}= u^2w+uw^2$ and
$e_3^{bc}= vw+v^2w^2$ and $e_4^{bc}= v^2w+vw^2$.

For  $i,j,k\in \{1,2,3\}$ and pairwise different, we have
\begin{eqnarray*}
2\cdot \F_{2^{\frac{(p_i-1)(p_j-1)}{2}}} & \cong &  \F_2G\cdot (1-\widehat{C_{p_i}})(1-\widehat{C_{p_j}})
\widehat{C_{p_k}}\\ 
& = &\,\F_2G\cdot f_1^{ij}\oplus \F_2G\cdot f_2^{ij},
\end{eqnarray*}
where

$f_1^{ij}=(1-\widehat{C_{p_i}})(1-\widehat{C_{p_j}})\widehat{C_{p_k}}\,+\,u_iv_j\widehat{C_{p_k}}\,+\,u_i^2 v_j^2\widehat{C_{p_k}}$
\\
and

$f_2^{ij}=(1-\widehat{C_{p_i}})(1-\widehat{C_{p_j}})\widehat{C_{p_k}}\,+\,u_iv_j^2\widehat{C_{p_k}}\,+\,u_i^2 v_j\widehat{C_{p_k}},$
\\
with
$u_i$ expressed as in~(\ref{uf2p}) in terms of the generator of $C_{p_i}$ and $v_j$ expressed as
in~(\ref{vf2p}) in terms of the generator of $C_{p_j}$.

Now we calculate the four simple components of $$\F_2C_{p_1}(1-\hat{a})\otimes \F_2C_{p_2}(1-\hat{b})\otimes \F_2C_{p_3}(1-\hat{c}).$$
In $\F_2C_{p_1}(1-\hat{a})\otimes \F_2C_{p_2}(1-\hat{b})$, we have $(uv)^3=(1-\hat{a})(1-\hat{b})$ and $uv\neq e_3^{ab}$.
Now take $\alpha=uve_3^{ab}=uv+u^2v+uv^2$.
Similarly, $(u^2v^2)^3=(1-\hat{a})(1-\hat{b})$ and $\alpha^2=u^2v^2e_3^{ab}=u^2v^2+u^2v+uv^2$.

Hence the elements $A=\alpha w+\alpha^2w^2$ and $B=\alpha^2 w+\alpha w^2$ are the primitive idempotents of
$(\F_2 [C_{p_1}\times C_{p_2}]) e_3^{ab}\otimes\F_2C_{p_3}(1-\hat{c})$

Similarly, in $\F_2C_{p_1}(1-\hat{a})\otimes \F_2C_{p_2}(1-\hat{b})$, we have $(uv^2)^3=(1-\hat{a})(1-\hat{b})$  and $uv^2\neq e_4^{ab}$.

Set $\beta=uv^2e_4^{ab}=uv^2+u^2v^2+uv$.

The elements $C=\beta w+\beta^2w^2$ and $D=\beta^2 w+\beta w^2$ are the primitive idempotents of
$(\F_2 [C_{p_1}\times C_{p_2}]) e_4^{ab}\otimes\F_2C_{p_3}(1-\hat{c})$.

Finally, it is an easy computation to show that $A,B,C$ and $D$ are orthogonal idempotents and
that $A+B+C+D=(1-\hat{a})(1-\hat{b})(1-\hat{c})$. Notice that:\\

$A=\alpha w+\alpha^2w^2=(uv+u^2v+uv^2)w+(uv+u^2v+uv^2)^2w^2=uvw+u^2vw+uv^2w+u^2v^2w^2+uv^2w^2+u^2vw^2=
(1-\hat{a})(1-\hat{b})(1-\hat{c})+ u^2v^2w+uvw^2$.\\

$B=\alpha^2 w+\alpha w^2=(uv+u^2v+uv^2)^2w+(uv+u^2v+uv^2)w^2= u^2v^2w+uv^2w+u^2vw+uvw^2+u^2vw^2+uv^2w^2=
(1-\hat{a})(1-\hat{b})(1-\hat{c})+ u^2v^2w^2+uvw$.\\

$C=\beta w+\beta^2w^2=(uv^2+u^2v^2+uv)w+(uv^2+u^2v^2+uv)^2w^2=uv^2w+u^2v^2w+uvw+u^2vw^2+uvw^2+u^2v^2w^2=
(1-\hat{a})(1-\hat{b})(1-\hat{c})+ u^2vw^2+uv^2w^2$.\\

$D=\beta^2 w+\beta w^2=(uv^2+u^2v^2+uv)^2w+(uv^2+u^2v^2+uv)w^2=u^2vw+uvw+u^2v^2w+uv^2w^2+u^2v^2w^2+uvw^2=
(1-\hat{a})(1-\hat{b})(1-\hat{c})+ uv^2w+u^2vw^2$.

\end{IEEEproof}

\section{Codes in $\F_2 (C_{p^m}\times C_{q^n})$, $m\geq 2, n\geq 2$}

The results in Section\ref{idemp} allow us to obtain the following.

\begin{theo}\label{theocpmcqn}
Let $p$ and $q$ satisfy~(\ref{hypopq}).
Let $G=\left< a \right>\times \left< b \right>$, with
$C_{p^m} = \left< a \right>$ and $C_{q^n}=\left< b \right>$.
Then the minimal codes of $\F_2(C_{p^m}\times C_{q^n})$ are described in the following table.
$$
\begin{array}{|l|c|c|c|} \hline \label{cpmcqn}
\mbox{\rm Ideal} & \mbox{\rm Primitive} & \mbox{\rm Dimension} & \mbox{\rm Code}  \\
             & \mbox{\rm Idempotent} &                 & \mbox{\rm Weight}   \\ \hline\hline
I_0 & \wh{a}\wh{b}       &  1      & p^mq^n  \\ \hline
I_{0j} & \wh{a}(\wh{b^{q^j}} + \wh{b^{q^{j-1}}})       &  q^{j-1}(q-1)      & 2p^mq^{n-j}  \\ \hline
I_{i0} & (\wh{a^{p^i}} + \wh{a^{p^{i-1}}}) \wh{b}       & p^{i-1}(p-1)      & 2p^{m-i}q^n  \\ \hline

I_{ij}^* & uv + u^2v^2   &  \frac{(p^{i}-p^{i-1})q^{j-1}(q-1)}{2}  & 
                \\
I_{ij}^{**} & uv^2 + u^2v       &  \frac{p^{i-1}(p-1)q^{j-1}(q-1)}{2}  & 
                 \\
                \hline
    \hline
\end{array}
$$
where 
\begin{eqnarray*}
u & = & \wh{a^{p^i}}(a^{2^0p^{i-1}}+a^{2^2p^{i-1}} + \cdots + a^{2^{p-3}p^{i-1}}), \\
                &  \mbox{if} & \,p\equiv 1(\!\!\!\!\!\mod 4)\;\;\mbox{or}\\
 u & = & \wh{a^{p^i}}(1+a^{2^0p^{i-1}}+a^{2^2p^{i-1}} + \cdots + a^{2^{p-3}p^{i-1}}),\\
                         & \mbox{if}& \,p\equiv 3(\!\!\!\!\!\mod 4) 
                        \end{eqnarray*}
 and
 \begin{eqnarray*}
  v & =& \wh{b^{q^j}}(b^{2^0q^{j-1}}+b^{2^2q^{j-1}} + \cdots + b^{2^{q-3}q^{j-1}}), \\
                        & \mbox{if}&\,q\equiv 1(\!\!\!\!\!\mod 4) \;\;\mbox{or}\\ 
 v & = & \wh{b^{q^j}}(1+b^{2^0q^{j-1}}+b^{2^2q^{j-1}} + \cdots + b^{2^{q-3}q^{j-1}}), \\
                         & \mbox{if}& \,q\equiv 3(\!\!\!\!\!\mod 4)  
                            \end{eqnarray*}                                                      
\end{theo}
\begin{IEEEproof}
Since $\bar{2}$ generates $U(\zzz_{p^2})$  by~\cite[Lemma~5]{FM}, we have
\begin{eqnarray*}
\F_2C_{p^m} & = &  \F_2C_{p^m}\cdot \wh{a}\oplus\bigoplus_{i=1}^{m} \F_2C_{p^m}\cdot (\wh{a^{p^i}} + \wh{a^{p^{i-1}}}) \\ & \cong &
\F_2 \oplus \bigoplus_{i=1}^{m} \F_{2^{(p^{i}-p^{i-1})}}, \\
\F_2C_{q^n} & = &  \F_2C_{q^n}\cdot \wh{b}\oplus\bigoplus_{j=1}^{n} \F_2C_{q^n}\cdot (\wh{b^{q^j}} + \wh{b^{q^{j-1}}}) \\ & \cong &
\F_2 \oplus \bigoplus_{j=1}^{n} \F_{2^{(q^{j}-q^{j-1})}}.
\end{eqnarray*}

Notice that since $\bar{2}$ generates $U(\zzz_{p^2})$ also $\bar{2}$ generates $U(\zzz_{p^m})$, 
hence, using~\eqref{hypopq} and Lemma~\ref{lemars}, we have the following decomposition:\\

\noindent
$\F_2(C_{p^m}\times C_{q^n})   = \F_2C_{p^m}\otimes_{\F_2} \F_2C_{q^n} $
\begin{eqnarray} 
 & \cong &
(\F_2 \oplus \bigoplus_{i=1}^{m} \F_{2^{(p^{i}-p^{i-1})}}) \otimes_{\F_2} (\F_2 \oplus \bigoplus_{j=1}^{n} \F_{2^{(q^{j}-q^{j-1})}}) \nonumber \\
 & =  & \F_2 \oplus \bigoplus_{i=1}^{m} \F_{2^{(p^{i}-p^{i-1})}} \oplus \bigoplus_{j=1}^{n} \F_{2^{(q^{j}-q^{j-1})}} \label{dimenscpmcqn}\\ 
 & \oplus &
2\cdot \bigoplus_{i,j} \F_{2^{\frac{(p^i - p^{i-1})(q^j - q^{j-1})}{2}}}.\nonumber 
\end{eqnarray}


For each pair $i,j$, the idempotent $e_{ij}=(\wh{a^{p^i}} + \wh{a^{p^{i-1}}})\cdot (\wh{b^{q^j}} + \wh{b^{q^{j-1}}})$
is not primitive and, by Lemma~\ref{lemae1e2}, it decomposes as sum of two primitive idempotents, namely $e_{ij}^{(1)}= uv+u^2v^2$  and $e_{ij}^{(2)}= uv^2+u^2v$,
where $u$ and $v$ are as in the statement of the theorem.
Thus the minimal ideals $I_{ij}^*=\left< uv+u^2v^2\right>$ and $I_{ij}^{**}=\left< u^2v+uv^2\right>$ are such
that $I_{ij}^*\oplus I_{ij}^{**}=\left< e_{ij}\right>$.

The dimension of each code follows from~\eqref{dimenscpmcqn}. 

\

Consider the code $I_{0j}=\left< e=\wh{a}(\wh{b^{q^j}} + \wh{b^{q^{j-1}}}) \right>$. The element $(b^{q^j}+b^{q^{j-1}})e=(1+ b^{q^{j-1}})\wh{a}\wh{b^{q^{j}}}\in I_{0j}$ has weight $2p^mq^{n-j}$. Since $\wh{b^{q^{j}}}=\wh{b^{q^{j-1}}}+ (\wh{b^{q^{j}}}+\wh{b^{q^{j-1}}})$, we have 
$$
(\F_2G)\wh{a}\wh{b^{q^{j}}} = (\F_2G)\wh{a}\wh{b^{q^{j-1}}}\oplus (\F_2G)e,
$$ 
hence $I_{0j}\subset (\F_2G)\wh{a}\wh{b^{q^{j}}}$.

\ 

 An element of $(\F_2G)\wh{a}\wh{b^{q^{j}}}$ is of the form
$(\sum_{i,k}\lambda_{ik}a^ib^k)\wh{a}\wh{b^{q^{j}}}=\sum_{i,k}\lambda_{ik}b^k\wh{a}\wh{b^{q^{j}}}$ and has weight $\ell p^mq^{n-j}$, 
as $supp(b^k\wh{a}\wh{b^{q^{j}}})\cap supp(b^t\wh{a}\wh{b^{q^{j}}})=\emptyset$ or $supp(b^k\wh{a}\wh{b^{q^{j}}})= supp(b^t\wh{a}\wh{b^{q^{j}}})$, for $0\leq k\neq t\leq q^n-1$.

\

Since $b^k\wh{a}\wh{b^{q^{j}}}\cdot e=b^k\wh{a}\wh{b^{q^{j}}} \wh{a}(\wh{b^{q^j}} + \wh{b^{q^{j-1}}}) =0$, we have $b^k\wh{a}\wh{b^{q^{j}}}\notin I_{0j}$.
Therefore, $\omega(I_{oj})=2p^mq^{j-n}$.
Similarly, we have $\omega(I_{i0})=2p^{m-i}q^n$. 

\

Let $1\leq i\leq m$ and $1\leq j\leq n$. To compute
the weight of the code $I_{ij}^*$ (and similarly for $I_{ij}^{**}$), we set
 $$H=\left< a^{p^{i-1}}\right>\times \left< b^{q^{j-1}}\right> \mbox{  and  } K=\left< a^{p^{i}}\right>\times \left< b^{q^{j}}\right>,$$ so
$K\leq H\leq G$ and $\frac{H}{K}\cong C_p\times C_q = \left< a^{p^{i-1}} K\right>\times \left< b^{q^{j-1}} K \right>$. 

\

Consider the isomorphism
$$
\psi :  (\F_2 H)\wh{K}    \longrightarrow  \F_2\left(\frac{H}{K}\right) $$
such that $a^{p^{i-1}}\wh{K}  \longmapsto    a^{p^{i-1}} K$ and $ b^{q^{j-1}}\wh{K}  \longmapsto   b^{q^{j-1}} K .$
Since $a^{p^{i}}\wh{K}=b^{q^{j}}\wh{K}=\wh{K}$, an element $\alpha\in (\F_2H)\wh{K}$ is of the form
\begin{equation}\label{alfaemF2H}
\alpha=\sum_{t=0}^{p-1}\sum_{\ell=0}^{q-1} \alpha_{t\ell} a^{tp^{i-1}} b^{\ell q^{j-1}}\wh{K}
\end{equation}

\

For $1\leq t_1,t_2\leq p-1$, $1\leq\ell_1,\ell_2\leq q-1$ and $(t_1,\ell_1)\neq (t_2,\ell_2)$, we have
$$t_1p^{i-1}\not\equiv t_2p^{i-1} \!\mod p^{i} \ \mbox{ or } \ \ell_1 q^{j-1}\not\equiv\ell_2 q^{j-1}\!\mod q^j,$$
 hence
$supp(a^{t_1p^{i-1}} b^{\ell_1 q^{j-1}}\wh{K})\cap supp(a^{t_2p^{i-1}} b^{\ell_2 q^{j-1}}\wh{K})=\emptyset.$
So
\begin{eqnarray*}
\omega({\alpha}) & = & \omega\left(\sum_{t=0}^{p-1}\sum_{\ell=0}^{q-1} \alpha_{t\ell} a^{tp^{i-1}} b^{\ell q^{j-1}}\wh{K}\right) \\
& = & \sum_{t=0}^{p-1}\sum_{\ell=0}^{q-1} \alpha_{t\ell} \omega\left( a^{tp^{i-1}} b^{\ell q^{j-1}}\wh{K} \right)\\
& = & \left(\sum_{t=0}^{p-1}\sum_{\ell=0}^{q-1} \alpha_{t\ell}\right) \omega(\wh{K}),
\end{eqnarray*}
where $\sum_{t=0}^{p-1}\sum_{\ell=0}^{q-1} \alpha_{t\ell}=\omega(\psi(\alpha))$.

\

Notice that $e=uv+u^2v^2=\wh{K} f$, with $f\in \F_2H$.
Considering $(\F_2H)e \subset (\F_2 G)e=I_{ij}^*$, take $$\beta=\sum_{\tilde{\mu}=0}^{p^m-1}\sum_{\tilde{\lambda}=0}^{q^n-1} \alpha_{\tilde{\mu}\tilde{\lambda}}a^{\tilde{\mu}}b^{\tilde{\lambda}}e\in
(\F_2G)e.$$  
Since  $a^{kp^{i-1}}b^\ell q^{j-1} e\in (\F_2H)e$, we may write
$$\beta=\sum_{\mu=0}^{p^{i-1}-1}\sum_{\lambda=0}^{q^{j-1}-1} \delta_{\mu\lambda}a^{\mu}b^{\lambda},$$
where 
$\delta_{\mu\lambda}\in(\F_2H)e$.

\

For $0\leq \mu_1,\mu_2\leq p^{i-1}-1$, $0\leq\lambda_1,\lambda_2\leq q^{j-1}-1$ and $(\mu_1,\lambda_1)\neq (\mu_2,\lambda_2)$, we have
$supp(\gamma_1a^{\mu_1} b^{\lambda_1})\cap supp(\gamma_2 a^{\mu_2} b^{\lambda_2})=\emptyset$, where $\gamma_1,\gamma_2\in(\F_2H)e$. 
Indeed, note that the exponents of $a$ and $b$ which appear in $\gamma_1a^{\mu_1} b^{\lambda_1}$ and $\gamma_2a^{\mu_2} b^{\lambda_2}$
are, respectively, $\mu_1+t_1p^{i-1}$, $\lambda_1+s_1q^{j-1}$ and $\mu_2+t_2p^{i-1}$, $\lambda_2+s_2q^{j-1}$. If 
$supp(\gamma_1a^{\mu_1} b^{\lambda_1})\cap supp(\gamma_2 a^{\mu_2} b^{\lambda_2})\neq\emptyset$, we should have 
$\mu_1+t_1p^{i-1}\equiv \mu_2+t_2p^{i-1} \!\mod p^m$ and $\lambda_1+s_1q^{j-1}\equiv\lambda_2+s_2q^{j-1}\!\mod q^n$, but this does not occur.

Hence,
$$\omega(\beta) =\omega\left( \sum_{\mu=0}^{p^{i-1}-1}\sum_{\lambda=0}^{q^{j-1}-1} \delta_{\mu\lambda}a^{\mu}b^{\lambda}\right)
=\sum_{\mu=0}^{p^{i-1}-1}\sum_{\lambda=0}^{q^{j-1}-1} \omega(\delta_{\mu\lambda}).$$
Thus, given a non-zero element $\beta\in(\F_2G)e$, there exists
a non-zero element in $\beta^{\prime}\in(\F_2H)e$ such that $\omega(\beta)\geq\omega(\beta^{\prime})$. 
Thus, $\omega((\F_2H)e)=\omega((\F_2G)e)$.

\end{IEEEproof}

\begin{example}\label{ex35} \end{example}

For $p=3$ and $q=5$, let $G=C_{3^m}\times C_{5^n}=\langle a\rangle\times \langle b\rangle$, with $o(a)=3^m$ and $o(b)=5^n$.
According to Theorem~\ref{theocpmcqn}, in $\F_2 (C_{3^m}\times C_{5^n})$ with
$1\leq i\leq m-1$ and $1\leq j\leq n-1$, the code $I_{ij}^*=\left< uv+u^2v^2\right>$
is generated by the element 

\begin{eqnarray*} 
e_{ij}^{(1)} & = &
 uv+u^2v^2 \\ & = & \wh{a^{3^{i+1}}}\wh{b^{5^{j+1}}}[(b^{5^j}+b^{2\cdot 5^j}+b^{2^2\cdot 5^j}+b^{2^3\cdot 5^j}) \\ 
 & + & a^{3^i}(b^{5^j}+b^{2^2\cdot 5^j})+ a^{2\cdot 3^i}(b^{2\cdot 5^j}+b^{2^3\cdot 5^j})]
 \end{eqnarray*}

Using Example~\ref{ex35} and the computations above, we get $$\omega(e_{ij}^{(1)})\,=\,\omega(I_{ij}^*)\,=\,3^{m-(i+1)}\cdot 5^{n-(j+1)}\cdot 8.$$

\ifCLASSOPTIONcaptionsoff
  \newpage
\fi

\end{document}